\documentclass{elsart}
\usepackage{graphicx}
\usepackage{amssymb}

\begin{document}

\begin{frontmatter}

\title{Decay to the nonequilibrium steady state of the thermal diffusion 
in a tilted periodic potential}

\author[waseda]{T.Monnai},
\ead{monnai@suou.waseda.jp}
\author[ichidai]{A.Sugita},
\ead{sugita@a-phys.eng.osaka-cu.ac.jp}
\author[ichidai]{J.Hirashima}, 
and
\author[ichidai]{K.Nakamura}
\ead{nakamura@a-phys.eng.osaka-cu.ac.jp}
\address[waseda]{Department of Applied Physics ,Waseda University, 3-4-1 Okubo,
Shinjuku-ku,Tokyo 169-8555,Japan}
\address[ichidai]{Department of Applied Physics, Osaka City University, 
3-3-138 Sugimoto, Sumiyoshi-ku, Osaka 558-8585, Japan}

\begin{abstract} 
We investigate asymptotic decay phenomenon towards the nonequilibrium 
steady state of the thermal diffusion in the presence of a tilted periodic 
potential.
The parameter dependence of the decay rate is 
revealed by investigating 
the Fokker-Planck (FP) equation in 
the low temperature case under the spatially periodic boundary condition (PBC).
We apply the WKB method to the associated
Schr\"odinger equation. 
While eigenvalues of the non-Hermitian FP operator are complex in general,
in a small tilting case
the imaginary parts of the eigenvalues are almost vanishing.
Then the Schr\"odinger equation is solved with PBC. 
The decay rate is analyzed
in the context of quantum tunneling through a triple-well effective periodic potential.
In a large tilting case, the imaginary parts of the eigenvalues of FP operator 
are crucial. We apply the complex-valued WKB method to 
the Schr\"odinger equation
with the absorbing boundary condition, finding that 
the decay rate saturates and depends only
on the temperature, the potential periodicity and the viscous constant. 
The intermediate tilting case is also explored. The analytic results 
agree well with the
numerical data for a wide range of tilting.
\end{abstract}

\end{frontmatter}

\section{Introduction}
Thermal diffusion process in a periodic potential in the presence of external force appears in many 
situations.  
In solid-state physics, it appears in diffusion of ions or molecules on crystal surfaces\cite{Frenken}, Josephson 
junctions\cite{Barone}, motion of fluxons in superconductors\cite{Shapiro}, 
rotation of molecules in solids\cite{Georgievskii}, superionic conductors\cite{Fulde}, charge 
density waves\cite{Gruner}, to mention a few.
As a ratchet system, it is also applied for
particle selection\cite{Ajdari}, biophysical processes\cite{Hoppenstead} and intracellular 
transport\cite{Reimann1}\cite{Hangi}.
Most of the researches in this context are concerned with the steady state.
For example, it is shown that a symmetric dichotomous noise causes non-zero steady state current in such a system\cite{Doering}.
The existence of an optimal diffusion coefficient vs. external force is also reported\cite{Reimann2}.

Such a diffusion process can be described by the over-damped Langevin equation with 
a tilted periodic potential,  
\begin{eqnarray}
&&\eta \dot{x}(t)=-\frac{\partial}{\partial x}U^{(0)} (x)+ \frac{2\pi W}{L} x +\xi(t),  \nonumber \\
&&\langle \xi(t)\rangle=0,\;\;\; \langle\xi(t)\xi(t')\rangle =2\eta\theta\delta(t-t').
\end{eqnarray}
Here, $\eta$, $\theta=k_{B}T$ and $W$ are the viscous coefficient, temperature and  
the external force, respectively.
$\xi(t)$ is the Langevin force which satisfies the fluctuation dissipation theorem.
The periodic potential $U^{(0)}(x+L)=U^{(0)}(x)$ with period $L$ is tilted by 
the external tilting force 
$F=\frac{2\pi W}{L}$ and we
call the potential $U(x)\equiv U^{(0)}(x)-\frac{2\pi W}{L} x$ as a tilted potential.

In order to investigate the probability density $P(x,t)$ that the particle
is found in a position $x$ at time $t$, it is useful to transform the Langevin equation into the 
corresponding
Fokker-Planck(FP) equation,
\begin{equation}
\frac{\partial}{\partial t}P(x,t)=\frac{1}{\eta}\frac{\partial}{\partial x}
\left(\frac{\partial 
U(x)}{\partial x}
+\theta\frac{\partial}{\partial x}\right)P(x,t).
\label{Fokker}
\end{equation}

Hereafter we solve FP equation (\ref{Fokker}) 
under the (spatially) periodic boundary condition (PBC), $P(x+L,t)=P(x,t)$, 
which is satisfied by the steady state at $t\rightarrow \infty$.
The analytic expression for the steady state 
$P^{st}(x)\equiv \lim_{t\rightarrow \infty}P(x,t)$ 
under PBC is well-known\cite{Reimann1,Risken,Lebowitz}:
\begin{eqnarray}
&&P^{st}(x)=N\frac{\eta}{\theta}\, e^{-\frac{U(x)}{\theta}}
\int_x^{x+L}dye^{\frac{U(y)}{\theta}},\nonumber 
\\
&&N=\frac{\theta}{\eta}
\left(\int_0^L dx\int_x^{x+L} dy e^{\frac{U(y)-U(x)}{\theta}}\right)^{-1},
\label{steady}
\end{eqnarray}
where $N$ is the normalization constant chosen so that $\int_0^Ldx P^{st}(x)=1$.
The steady state has a non-zero probability current $J^{st}(x)$ given by
\begin{equation}
J^{st}\equiv \frac{1}{\eta}
\left(U(x)-\theta\frac{\partial}{\partial x}\right)P^{st}(x)=
N\left(1-e^{-\frac{2\pi 
W}{\theta}}\right).
\end{equation}

The above steady state is a nonequilibrium state.
In this paper, we study the low-temperature relaxation dynamics  
towards the steady state in the presence 
of a tilted periodic potential. In particular, asymptotic behaviors
of the decay phenomenon is investigated 
by the eigenfunction expansion method. Our main interest lies in parameter
dependence of the decay rate. 

This paper is organized as follows.
In section \ref{eigenvalues} we present the framework of 
our analysis, relevant numerical data for eigenvalues of the FP operator, 
and the main result.
Section \ref{WKBsec} treats the important theme on the boundary condition
for eigenstates of
the associated Schr\"odinger equation. 
In subsection \ref{small_tilting} the small tilting case 
is investigated in the context of a tunneling problem.
In subsection \ref{large_tilting} 
the large tilting case is treated, where we shall apply the complex-valued
WKB method to the Schr\"odinger equation with the absorbing boundary 
condition deduced from PBC for the original FP equation. 
Section \ref{summary} is devoted to a summary. 
Appendix is concerned with detailed calculations
in section \ref{large_tilting}.

\section{Eigenvalues of the Fokker-Planck operator}
\label{eigenvalues}
The WKB treatment of FP equation has a long history since the monumental work by van 
Kampen\cite{vanKampen}.
Most of the studies, however, were limited to 
the system with a well-defined thermal 
equilibrium
state\cite{Caroli}. This fact guaranteed a real-valued nature 
of eigenvalues for the non-Hermitian FP operator, 
and one can reduce  
FP equation to the eigenvalue problem of the associated Schr\"odinger equation  
with the same natural boundary condition as the FP equation has.
On the other hand, relaxation dynamics towards the nonequilibrium steady state with
the
non-vanishing constant current
demands a qualitatively different WKB approach.
In fact FP operator in such a system has complex  
eigenvalues in general. 

This situation is recognized in our system with a tilted sinusoidal potential\cite{Noziere},
\begin{equation}
U(x)=U_0 \cos\left(\frac{2\pi x}{L}\right)-\frac{2\pi W}{L}x ,
\label{orignal}
\end{equation}
where $U_0$ is the amplitude of the periodic part of the potential.
Putting eigenvalues of the non-Hermitian FP operator 
as $-E$ and substituting $P(x,t)=e^{-Et}P(x)$ 
into Eq.(\ref{Fokker}), we have
the eigenvalue problem, 
\begin{equation}
-E P(x)=\frac{1}{\eta}\frac{\partial}{\partial x}
\left(\frac{\partial U(x)}{\partial x}
+\theta\frac{\partial}{\partial x}\right)P(x).
\label{Eigenvalue}
\end{equation}

While the steady state in Eq.(\ref{steady}) is 
periodic, $P^{st}(x+L)=P^{st}(x)$, and is characterized by 
the zero Bloch wavenumber ($k=0$), it is difficult to
analyze the relaxation dynamics in general.  Any initial distribution 
$P(x,0)$ which include
$k\neq 0$ Fourier components can relax towards $P^{st}(x)$: 
the $k\neq 0$ components will decay out on the way of time evolution.
In the present work, however,  we confine ourselves to the manifold of $k=0$,
which makes the analysis of the decay phenomenon accessible.
This implies that
the time-dependent
solution of FP equation in Eq.(\ref{Fokker})
relaxing towards the steady state should
satisfy the spatial PBC, $P(x+L,t)=P(x,t)$
and be constructed from solutions 
with a natural boundary condition 
$Q(x,t)$ ($\lim_{x\rightarrow \pm \infty}Q(x,t)=0$)
as 
\begin{equation}
P(x,t)=\sum_{n=-\infty}^{\infty}Q(x+nL,t).
\label{solution}
\end{equation}
Since the FP operator in Eq.(\ref{Fokker}) is invariant against 
space translation by $L$, 
it has Bloch-type eigenstates 
$\{P_{n,k}(x)\}$ and (complex) eigenvalues $\{E _{n,k}\}$.
Noting the normalization matrix
$\int_0^L dxP_{k,n}^*(x) P_{k',n'}(x) = \delta_{k,k'} N_{n,n'}^{(k)}$
due to the non-Hermitian nature of FP operator,
the solution
$P(x,t)$ in Eq.(\ref{solution}) is now written explicitly as
\begin{eqnarray}
&&P(x,t) = \sum_{n}C_{n,k=0}P_{n,k=0}(x)
e^{-E _{n,k=0} t},\label{wp} \\
&&C_{n,k=0} =\sum _{n'}(N^{(0)})^{-1}_{n,n'} \int_0^L dxP_{n',k=0}^*(x)P(x,0),
\qquad  C_{n,k\neq 0} =0,
\end{eqnarray}
which starts from an arbitrary $k=0$ distribution $P(x,0)$ at $t=0$ and relaxes
towards $P^{st}(x)$.


In the manifold with $k=0$, the probability distribution
can be expanded as
$P(x)=\sum_{n}c_n e^{\frac{2\pi i x}{L}n}$. 
Then the eigenvalue problem (\ref{Eigenvalue}) is reduced to 
\begin{eqnarray}
&&\left(\frac{2\pi}{L}\right)^2\sum_m A_{n,m} c_m=\eta E c_n\nonumber \\
&&A_{n,m}=-\left(inW+\theta n^2\right)\delta_{n,m}-\frac{n U_0}{2}\delta_{n-1,m}+\frac{nU_0}{2}\delta_{n+1,m},
\label{complex}
\end{eqnarray}
which can be solved numerically by truncating the infinite dimensional
matrix $A$ to a large but finite dimendional matrix.

\begin{figure}
\center{
\includegraphics[scale=0.8]{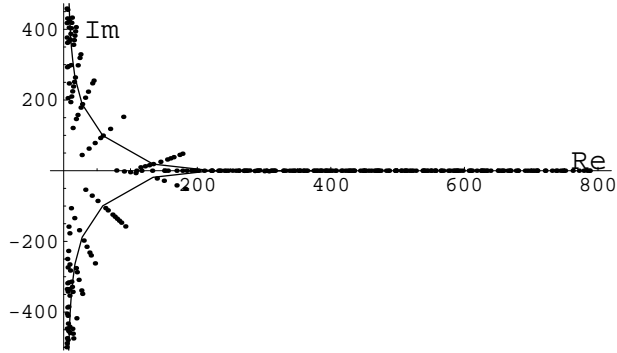}
}
\caption{Distribution of the complex eigenvalue with the second smallest real part
for full parameter range $3\leq U_0\leq 20, 0\leq W\leq 20$ (dots), 
and for the case of fixed potential barrier $U_0=10, 0\leq W\leq 20$ (solid line).
A pair of solid lines rapidly merge to the real axis as $W$ decreases below $U_0$.}
\label{Figg1}
\end{figure}

The zero-eigenvalue ($E=0$)
corresponds to the steady state.
Under PBC, as is obvious from Eq. (\ref{wp}), 
the probability distribution approaches exponentially to
the unique steady state, and the decay rate $\lambda$ is 
obtained as the real part of the second smallest eigenvalue. 
Among many complex eigenvalues of Eq.(\ref{complex}), therefore,
we have focused on the one 
with the second-smallest real part and its variation 
against various parameter values $U_0, W$ (see Fig.\ref{Figg1}).
From Fig.\ref{Figg1} one can see that the eigenvalues 
with large positive real 
parts have nearly zero imaginary part. (See tails of solid lines in Fig. \ref{Figg1}.)
These  eigenvalues correspond to the small tilting case.
In fact, a pair of solid lines rapidly merge to the real axis as $W$ decreases below $U_0$,
ensuring the real-valued nature of eigenvalues in the case of small tilting.

\begin{figure}
\center{
\includegraphics[scale=0.8]{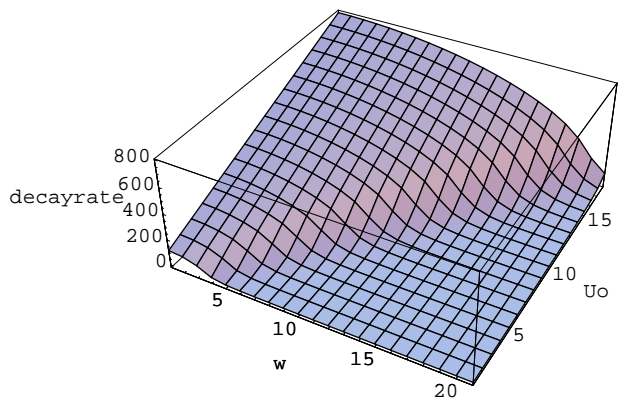}
}
\caption{$W$ and $U_0$ dependence of the decay rate $\lambda$.}
\label{Figg2}
\end{figure}

$U_0$ and $W$ dependence of the decay rate $\lambda $ is shown in Fig. \ref{Figg2},
which we have confirmed
by numerically solving the FP equation in Eq.(\ref{Fokker}).
There is a smooth but steep decrease of the decay rate around 
the crossover line of $U_0 \simeq W$, where the original potential minima disappear.

In the next section, the Schr\"odinger 
equation associated with FP equation
will be solved with use of the WKB method.  
In the small tilting case in section \ref{small_tilting}, 
we impose PBC for the wavefunction
of the Schr\"odinger 
equation and apply the Bloch theorem.
The decay rate for $W<U_0$ is shown to be
$\left(\frac{2\pi}{L}\right)^2\sqrt{1-\left(\frac{W}{U_0}\right)^2} U_0$. 
In the large tilting case in subsection \ref{large_tilting}, 
we have recourse to the absorbing boundary condition
and apply the WKB method which admits complex energies. 
The decay rate for $W\gg U_0$ is found to be 
$\left(2\pi/L\right)^2\theta $ (with no $W$ and $U_0$ dependence).
The crossover region where $W$ is 
comparable to $U_0$ is numerically analyzed based on this approach. 

\begin{figure}
\center{
\includegraphics[scale=0.8]{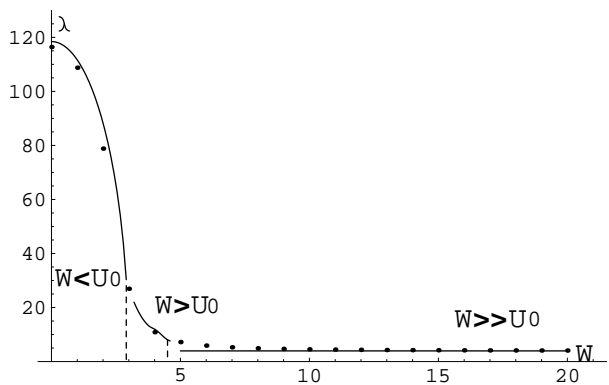}
}
\caption{$W$ dependence of the decay rate $\lambda$
for $U_0=3, \theta=0.1$. 
Dots show the decay rates obtained from numerical diagonalization
of FP operator. 
Three solid 
lines are 
theoretical results (For small tilting case $W<U_0$ 
the decay rate $\lambda$ is obtained as the first excited level
 $\hbar\omega_a$ of the associated Schr\"odinger 
equation (left part). For crossover regime $W\sim U_0$, $\lambda$ is given as the 
 numerical solution of (\ref{WKB2})(middle part). 
 For large tilting case $W\gg U_0$, the perturbative solution of 
(\ref{complexWKB}) gives 
$\lambda$ (right part).   }
\label{Figg3}
\end{figure}

\section{WKB analysis of probability distribution}
\label{WKBsec}

The FP equation is transformed into the Schr\"odinger equation by the separation 
ansatz\cite{vanKampen}
\begin{equation}
P(x,t)=e^{-\frac{U(x)}{2\theta}}\phi(x) e^{-E t}.
\label{separation}
\end{equation} 
Then one has 
\begin{equation}
\theta\frac{d^2}{dx^2}\phi(x)+(E-V(x))\phi(x)=0,
\label{Eigen2}
\end{equation}
where we redefined eigenvalue as $E\eta\rightarrow E$,
 and defined the effective potential 
\begin{eqnarray}
V(x)\equiv \theta\left\{\left(\frac{U^{'}(x)}{2\theta}\right)^2-
\frac{U^{''}(x)}{2\theta}\right\}.
\label{secondary}
\end{eqnarray}

The Schr\"odinger operator (SO) in Eq.(\ref{Eigen2})  looks Hermitian.
Noting in Eq.(\ref{separation}) $U(x+L)=U(x)-2\pi W$ 
and $P_{n,k}(x+L)=e^{ikL}P_{n,k}(x)$, however,
the wave function $\phi(x)$
of the Schr\"odinger equation should satisfy 
the absorbing boundary condition,
\begin{equation}
\phi(x+L)=e^{ikL-\frac{2\pi W}{2\theta}}\phi(x),
\label{absorb}
\end{equation}
which eventually breaks the Hermitian nature of SO.

In the following we shall analyze the Schr\"odinger equation in Eq.({\ref{Eigen2})
first in the small tilting case and then in the strong tilting case.
In the small tilting case, 
the numerical evidence in the previous section
has shown the real-valued nature of the second-lowest eigenvalue,
indicating the recovery of Hermitian nature of SO.
Therefore we reduce Eq. (\ref{absorb}) to
\begin{equation}
\phi(x+L)=e^{ikL}\phi(x),
\label{reduce}
\end{equation}
by assuming the absorbing term $\frac{2\pi W}{2\theta}$ being vanishing.
The validity of this assumption will be verified by comparing between the analytic and
numerical decay rates. Below we shall confine to the $k=0$ manifold.

\subsection{Small tilting case}
\label{small_tilting}

Substituting the tilted sinusoidal potential (\ref{orignal})
into (\ref{secondary}), one has
the effective potential,
\begin{equation}
V(x) = \frac{\pi^2}{L^2\theta}\left(U_0\sin\frac{2\pi x}{L}+W\right)^2
+\frac{1}{2}\left(\frac{2\pi}{L}\right)^2
U_0\cos\frac{2\pi}{L}x.\label{EffectivePotential}
\end{equation} 

Since the first term is much larger than the second one
in the low temperature regime $\theta\ll U_0$,
the second one is omitted hereafter. 

In solving Eq.(\ref{Eigen2}) the low temperature condition allows us to use 
the WKB approximation, since $\theta$ corresponds to 
$\frac{\hbar^2}{2m}$ in the standard Schr\"odinger equation.  
In the case that the external force is weak ($W<U_0$), 
the effective potential $V(x)$ has
three wells. 
We consider the energy levels for the eigenstates which 
experience all the three wells\cite{threewell} (see Fig. \ref{Figg4}).
\begin{figure}
\center{
\includegraphics[scale=0.8]{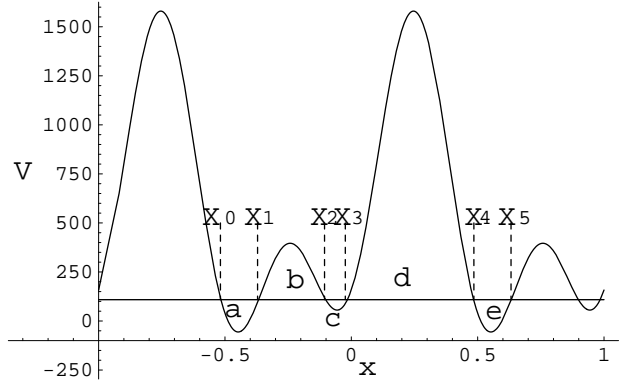}
}
\caption{Effective potential $V(x)$ and the real part of the eigenvalue of FP operator  
with second smallest real part $E$ with 
the corresponding classical turning points $x_0,....,x_5$ 
for $W=1,U_0=3$.}
\label{Figg4}
\end{figure}

In terms of the tilted potential $U(x)$, the weak external force guarantees the existence of potential 
minima in Fig. \ref{Figg5}.
\begin{figure}
\center{
\includegraphics[scale=0.8]{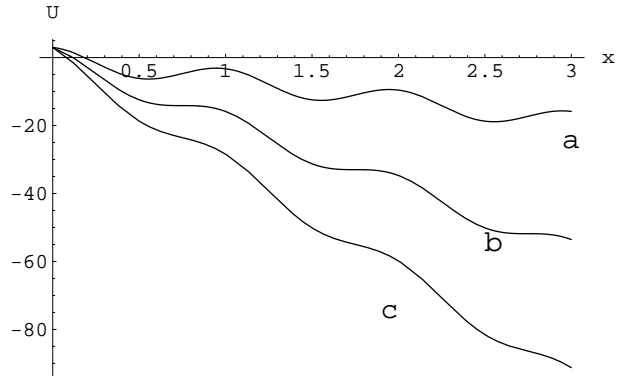}
}
\caption{Original tilted potential $U(x)$ in real space. Small tilting case (a), 
crossover region (b), and large tilting case (c).}
\label{Figg5}
\end{figure}
Let us denote the wave functions in each region $a,\cdot\cdot\cdot,e$
bordered by classical turning points $x_0,\cdot\cdot\cdot,x_5$
as $\phi_a,\cdot\cdot\cdot,\phi_e$ and introduce
the action integrals 
\begin{eqnarray}
&&S_a=\frac{1}{\hbar}\int_{x_0}^{x_1}p dx,\;\;\; 
M_b=\frac{1}{\hbar}\int_{x_1}^{x_2}pdx\nonumber \\
&&S_c=\frac{1}{\hbar}\int_{x_2}^{x_3} p dx,\;\;\;
M_d=\frac{1}{\hbar}\int_{x_3}^{x_4} p dx,
\end{eqnarray}
where $\hbar^2=2m\theta$ and $p=\sqrt{2m|E-V(x)|}$ 
($p$ stands for the momentum in a potential-well region). 
$S$ and $M$ are defined in each well and barrier, respectively.
Each neighboring wave functions are mutually related by the connection formula
\begin{equation}
\frac{1}{\sqrt{p}}e^{\pm i (S+\frac{\pi}{4})}\quad(E>V(x))\quad 
\longleftrightarrow 
\quad \frac{1}{\sqrt{p}}
\left(e^{S}\pm \frac{i}{2}e^{-S}\right)\quad(E<V(x)),
\end{equation} 
where
\begin{eqnarray} 
S=\mid \frac{1}{\hbar}\int_{x}^{y_0} p dx \mid,
\end{eqnarray}
with $y_0$ being the classical turning point.

For example, $\phi_a$ and $\phi_b$ are related as
\begin{eqnarray}
\phi_a(x)&=&\frac{1}{\sqrt{p}}
\left(c_1e^{-\frac{i}{\hbar}\int_{x_0}^x p dx}
+c_2e^{\frac{i}{\hbar}\int_{x_0}^x p dx}\right)\nonumber \\
\phi_b(x)&=&\frac{1}{\sqrt{|p|}}
\left\{c_1e^{-iS_a}\left(e^{\frac{1}{\hbar}\int_{x_1}^x p dx}
+ \frac{i}{2}e^{-\frac{1}{\hbar}\int_{x_1}^x p dx}\right)\right.\\
&&
\left.
+ c_2e^{iS_a}\left(e^{\frac{1}{\hbar}\int_{x_1}^x p dx}
- \frac{i}{2}e^{-\frac{1}{\hbar}\int_{x_1}^x p dx}\right)\right\}
\end{eqnarray}
In the same way, one obtains the expression for $\phi_e(x)$ as
\begin{equation}
\phi_e(x)=\frac{1}{\sqrt{p}}
\left\{(A c_1+B c_2)e^{-\frac{i}{\hbar}\int_{x_4}^x p dx}+
(C c_1+D c_2)e^{\frac{i}{\hbar}\int_{x_4}^x p dx}\right\},
\end{equation}
where $A$,$B$,$C$ and $D$ are given as
\begin{eqnarray}
A &=& 
e^{iS_a} \left\{
\frac{1}{2} \left(ie^{M_b}\sin S_c + \frac{1}{4}e^{-M_b}\cos S_c\right)e^{-M_d}
+ 2\left(e^{M_b}\cos S_c + \frac{i}{4}e^{-M_b}\sin S_c\right) e^{M_d} \right\},\nonumber \\
B &=&
e^{-iS_a} \left\{
\frac{1}{2}\left(e^{M_b}\sin S_c + \frac{i}{4}e^{-M_b}\cos S_c \right)e^{-M_d}
- 2\left(ie^{M_b}\cos S_c + \frac{1}{4}e^{-M_b}\sin S_c\right) e^{M_d}\right\},\nonumber \\
C &=& 
e^{iS_a}\left\{
\frac{1}{2}\left(e^{M_b}\sin S_c -\frac{i}{4}e^{-M_b}\cos S_c\right) e^{-M_d}
- 2\left(-ie^{M_b}\cos S_c + \frac{1}{4}e^{-M_b}\sin S_c\right) e^{M_d}\right\}\nonumber \\
D &=& 
e^{-iS_a}\left\{\frac{1}{2}\left(-ie^{M_b}\sin S_c + \frac{1}{4}e^{-M_b}\cos S_c \right)e^{-M_d}
+ 2\left(e^{M_b}\cos S_c -\frac{i}{4}e^{-M_b}\sin S_c\right) e^{M_d} \right\}\nonumber \\
&&
\end{eqnarray}

\indent On the other hand, with use of the Bloch theorem 
in Eq.(\ref{reduce}) based on the periodicity of 
the effective potential $V$,
$\phi_e$ and $\phi_a$ should be related as
\begin{equation}
\phi_e =e^{ikL}\phi_a,
\end{equation}
where $k$ is the wave number.
This relation gives a constraint to guarantee the nontrivial 
coefficients ($c_1, c_2$) of $\phi_a$ and $\phi_e$:
\begin{equation}
\left|
\begin{array}{cc}
A-e^{ikL} & B\\
C & D-e^{ikL}
\end{array} 
\right|=0,
\label{einen}
\end{equation}
which can be rewritten as 
\begin{equation}
e^{ikL}=\cos(kL)+ i\sin(kL)=\frac{1}{2}(A+D)\pm i\sqrt{1 - \frac{1}{4}(A+D)^2}.
\label{AD}
\end{equation}
Here we have used the fact $AD-BC=1$.
At low temperatures the tunneling 
integrals
are sufficiently large, $M_b\gg 1$ and $M_d\gg 1$. 
Then we can make the approximation:
\begin{equation}
\frac{1}{4}(A+D)^2\cong 4e^{2M_b}e^{2M_d}\cos^2(S_a)\, \cos^2(S_c) ,
\label{aproxAD}
\end{equation}
Since (\ref{aproxAD}) should satisfy (\ref{AD}), 
$|\cos S_a |$ and $|\cos S_c |$ must be sufficiently small and be
approximated as
\begin{eqnarray}
\cos(S_{a,c})&\simeq &(-1)^{n_{a,c}}
\left\{S_{a,c}-(n_{a,c}+\frac{1}{2})\right\}\pi\nonumber \\
\sin(S_{a,c})&\simeq &(-1)^{n_{a,c}}.
\end{eqnarray}
Furthermore, using the harmonic approximation in the neighborhood of the
potential minima 
$x_{min}^{a,c}$, the actions
can be evaluated as
\begin{equation}
S_{a,c}=\frac{\pi}{\hbar\omega_{a,c}}(E-V_{0\, a,c}),
\end{equation}
where $\omega_{a,c}=\sqrt{\frac{2\theta}{\hbar^2}V^{''}(x_{min}^{a,c})}$ is the curvature 
of the potential 
minima, and
$V_{0 a,c}$ stands for the local minima of $V(x)$. 
Then Eq.(\ref{AD})
gives the quantization condition for eigenvalues $E$
\begin{eqnarray}
&&\cos(kL)=\frac{1}{2}(A+D)\nonumber \\
&=&\frac{1}{2}\left\{
\left(4e^{M_b+M_d}+\frac{1}{4}e^{-M_b-M_d}\right)\cos(S_a)\cos(S_c)
-\left(e^{-M_b+M_d} + e^{M_b-M_d}\right)
\sin(S_a)\sin(S_c)\right\}\nonumber \\
&\simeq &
\frac{1}{2}\left\{
\left(4e^{M_b+M_d}+\frac{1}{4}e^{-M_b-M_d}\right)
\left((n_a+\frac{1}{2})\pi-\frac{\pi}{\hbar\omega_a}(E-V_{0 a})\right)
\left((n_c+\frac{1}{2})\pi-\frac{\pi}{\hbar\omega_c}(E-V_{0 c})\right)
\right.\nonumber \\
&&
\left.
-\left(e^{M_{b}-M_{d}}+e^{-M_b+M_d}\right)\right\}(-1)^{n_a+n_c}.
\label{qtcond}
\end{eqnarray}

Solving the quadratic equation of $E$ in (\ref{qtcond}), we obtain the lowest few
eigenvalues within the $k=0$ manifold,
\begin{eqnarray}
E_{\pm}&=&E_{0+}\pm \sqrt{E_{0-}^2+\epsilon}\nonumber \\ 
&\simeq&
E_{0+}\pm \left(E_{0-}+\frac{\epsilon}{2E_{0-}}\right),
\end{eqnarray}
where $E_{0\pm}$ stand for the
harmonic oscillator levels at the wells and $\epsilon$ is 
a correction due to tunneling, which are defined as
\begin{eqnarray}
E_{0\pm }
&=&
\frac{1}{2}\left\{
V_{0 a}\pm V_{0 c} + 
\hbar \omega_a \left(n_a+\frac{1}{2}\right)\pm \hbar\omega_c 
\left(n_c+\frac{1}{2}\right) \right\},
\nonumber \\
\epsilon\equiv\epsilon(k=0)
&=&
\frac{\hbar^2\omega_a\omega_c}{\pi^2}\frac{e^{-2M_b}+e^{-2M_d}+2(-1)^{n_a+n_c}e^{-M_b-M_d}
\cos k}
{4+\frac{1}{4}e^{-2M_b-2M_d}}.
\end{eqnarray}
In general, the eigenvalues form subbands with their width given by the tunneling term 
$\epsilon$, although we confine ourselves to the $k=0$ state in each of the subbands.
At $x_{min}^a=\frac{L}{2\pi}(n\pi - (-1)^n\arcsin\frac{W}{U_0})$, we find $V_{0a}=V(x_{min})\simeq -\frac{1}{2}\left(\frac{2\pi}{L}\right)^2\sqrt{1-\left(\frac{W}{U_0}\right)}U_0$ and
\begin{equation}
\hbar\omega_a\simeq \left(\frac{2\pi}{L}\right)^2
\sqrt{1-\left(\frac{W}{U_0}\right)^2}U_0.
\label{hw}
\end{equation}
Noting further that $V_{0a}+\frac{1}{2}\hbar\omega _a\simeq V_{0c}+\frac{1}{2}\hbar\omega _c\simeq 0$,
one can confirm the lowest eigenvalue $E_{-}\simeq 0$, 
which is responsible to the steady state.
Since the tunneling 
term is 
small compared to $\hbar\omega_a $, one finds the decay rate 
(:the second-lowest eigenvalue),
\begin{equation}
\lambda=E_{+}\simeq\hbar\omega_a= \left(\frac{2\pi}{L}\right)^2
\sqrt{1-\left(\frac{W}{U_0}\right)^2}U_0.\label{hw}
\end{equation}
We have confirmed numerically that this expression
gives a good approximation for the decay rate in the small tilting case (see Fig.\ref{Figg3}).

Now we can explain the $W$ (external force) dependence of the decay rate.
From (\ref{hw}) one has the monotonic $W$ (external force) dependence of the decay rate.
Intuitively, this parameter dependence of the decay rate is explained in terms of tunneling process as follows.
The energy level $\hbar\omega_a=\sqrt{2\theta V''(x_{min}^a)}$ is a 
monotonic function of the curvature of 
the potential minimum $V''(x_{min}^a)$ which decreases for increasing $W$, 
because the potential barrier becomes lower and
finally the barrier disappears as $W$ increases.
The case that each barrier separating adjacent local potential minima 
vanishes is treated in the next section. 

\subsection{Large tilting case}
\label{large_tilting}
The separation ansatz $P(x,t)=e^{-\frac{U(x)}{2\theta}}\phi(x) e^{-E t}$
in Eq.(\ref{separation})  seems to
contain a divergent term coming from $e^{\frac{2\pi W x}{2L\theta}}$ involved in
$e^{-\frac{U(x)}{2\theta}}$. This problem was overcome
by choosing the absorbing boundary condition in Eq.(\ref{absorb}). 
Confining ourselves to $k=0$ states as
in the previous subsection, we have
\begin{equation}
\phi(x+L)=e^{-\frac{2\pi W}{2\theta}}\phi(x).
\label{absorb2}
\end{equation}
With use of the complex-valued WKB method, Eq. (\ref{absorb2}) leads 
to the quantization condition for the complex eigenvalues $E=E_1+i E_2$ as
\begin{equation}
e^{\pm i\int_0^L dx \sqrt{\theta^{-1}(E_1-V(x)+iE_2)}}=e^{-\frac{2\pi W}{2\theta}+2n\pi i}
\label{WKB}
\end{equation}
with $V(x)$ given in Eq.(\ref{EffectivePotential}).

Remarkably there is no classical turning point on the real axis, 
and the energy becomes complex, as 
is confirmed numerically for the strongly tilted case (see Fig.\ref{Figg1}).
Then the decay rate $\lambda$ is given by the real part of the
eigenvalue with the second smallest part, which we denote $E_1$.
The existence of the imaginary part, which we denote $E_2$, 
implies that the asymptotic behavior is oscillating.

Noting again that the second term of (\ref{EffectivePotential}) is 
suppressed at low temperatures, 
the WKB quantization condition (\ref{WKB}) is rewritten as
\begin{eqnarray}
&&\frac{\pi W}{\theta L}\int_0^L\sqrt{1+2\frac{U_0}{W} 
\sin\frac{2\pi x}{L} + \left(\frac{U_0}{W}\right)^2\sin^2\frac{2\pi x}{L}-
\frac{\theta L^2}{\pi^2W^2}(E_1+iE_2)}\nonumber \\
&&=-\frac{\pi W}{\theta}+i 2\pi n .
\label{WKB2}
\end{eqnarray}
For the large tilting case $W\gg  U_0$, we can solve this equation perturbatively 
with the 
expansion parameter $U_0/W$ (see Appendix).
Since it is numerically shown that the integer $n$ is $\pm 1$ for 
the parameter range $3\leq U_0\leq 20,U_0\leq W\leq 20$, 
we investigate the case that $n=\pm 1$.
As a result, we obtain for large $W$
\begin{eqnarray}
E_1&=&\left(\frac{2\pi}{L}\right)^2\theta \nonumber \\
E_2&=&\pm \left(\frac{2\pi}{L}\right)^2 W.
\label{complexWKB}
\end{eqnarray}
These results are consistent with those from 
the numerical diagonalization of FP operator.
It is interesting that for large $W$, the decay rate $E_1$ saturates and depends only on 
the temperature, the period and the viscosity.
It is also worth noting that $E_2$, which represents
the oscillation frequency in the asymptotic behavior,  
is proportional to $W$ and does not depend on $U_0$.   

\section{Summary}
\label{summary}
The low-temperature ($\theta=k_{B}T$) decay of 
the thermal diffusion in the tilted periodic potential (period $L$, amplitude $U_0$ and
external tilting force $W$), which 
can be described by the Fokker-Planck (FP) equation, 
is investigated in terms of the (generalized) WKB method.
The nonequilibrium steady state demands a 
subtle approach to the asymptotic decay phenomenon, 
qualitatively different from the one
often used in the context of the decay to the thermal equilibrium.
However, we have confined ourselves to the manifold of $k=0$ which
includes the steady state. This recourse
makes the analysis of the decay phenomenon accessible.

While eigenvalues of FP operator are complex in general,
in a small tilting case ($W<U_0$)
the imaginary parts of the eigenvalues are shown to be vanishing.
Then the Schr\"odinger equation associated with FP equation
has the WKB solution satisfying PBC. 
Among the continuum of Bloch states, the zero-wavenumber eigenvalues 
are essential:
the lowest eigevalue is responsible for
the steady state and
the decay rate is determined
by the second-lowest eigenvalue within the zero-wavenumber manifold. 
The parameter dependence of the decay rate is  explained 
by the tunneling process through a triple-well effective periodic potential.
The decay rate $\lambda$ is given by 
$\left(\frac{2\pi}{L}\right)^2\sqrt{1-\left(\frac{W}{U_0}\right)^2} U_0$. 
In a large tilting case ($W\gg U_0$), 
the imaginary parts of the eigenvalues of FP operator 
are crucial. We apply the complex-valued WKB method to 
the Schr\"odinger equation
with the absorbing boundary condition, finding
the decay phenomenon characterized by the decay rate
$E_1 = \left(\frac{2\pi}{L}\right)^2\theta $ and the oscillation frequency
$E_2 = \left(\frac{2\pi}{L}\right)^2 W$.
The intermediate tilting case is also explored, and
our theoretical results from the WKB analysis 
explain $W$ and $U_0$ 
dependence of the numerical decay rate quite well for 
all parameter ranges in Fig.\ref{Figg3}.
The decay phenomena starting from
$k\neq 0$ distributions and
the contribution from continuous spectra to
nonexponential decays
will constitute subjects which we intend to study in future.

\section*{Acknowledgment}
T.M. thanks to Prof. S.Tasaki for valuable comments including the 
advice about calculation in Eq.(\ref{WKB}).
This work is partly supported by JSPS Research Fellowship for young scientist and
21st Century COE Program (Holistic research and Education, Culture,
Sports, Science and Technology). 
A.S. and K.N. are grateful to JSPS for the financial support to the Fundamental
Research.

\appendix
\section{Perturbative analysis
of the complex eigenvalue problem for large tilting case}
For the large tilting case ($W\gg U_0$), the WKB quantization condition
under the absorbing boundary condition is given by Eq.(\ref{WKB})
or equivalently by 
\begin{equation}
i\int_0^L dx \sqrt{\theta^{-1}(E_1-V(x)+iE_2)}=
-\frac{\pi W}{\theta}+2n\pi i . 
\label{absorbing}
\end{equation}
We have solved Eq.(\ref{absorbing}) perturbatively 
with use of the expansion parameter $\epsilon =\frac{U_0}{W}$ as follows.

We restrict ourselves to the case  $n=\pm 1$ which is observed numerically in the parameter range
$3\leq U_0\leq 20$, $U_0\leq W\leq 20$ and use a branch of square-root with positive imaginary part in the left
hand side of (\ref{absorbing}) (another 
branch does not satisfy (\ref{absorbing})). 
At first, we note that (\ref{absorbing}) is equivalent to 
\begin{eqnarray}
&&-\int_0^L dx\left(\sqrt{\frac{1}{2}(A+\sqrt{A^2+B^2}}) \nonumber
+i\sqrt{\frac{1}{2}(-A+\sqrt{A^2+B^2}})\right) \\
&=& -L+i\frac{2\theta L}{W}, \nonumber \\
A &=& \left(1+\epsilon \sin\frac{2\pi x}{L}\right)^2
 - \frac{\theta L^2}{\pi^2 W^2}E_1\nonumber \\
B &=& \frac{\theta L^2}{\pi^2 W^2}E_2.\label{perturbation}
\end{eqnarray}
We expand eigenvalue $E_1+iE_2$ up to the second order of $\epsilon$ as
\begin{equation}
\frac{\theta L^2}{\pi^2 W^2}E_1=A_1\epsilon+A_2\epsilon^2+O(\epsilon^3), \;\;\;
\frac{\theta L^2}{\pi^2 W^2}E_2=B_1\epsilon+B_2\epsilon^2+O(\epsilon^3).
\end{equation}
Then the real part of the equation (\ref{perturbation}) is expanded in $\epsilon$ as 
\begin{eqnarray}
&&\int_0^L dx\left\{1 + \frac{1}{2}(-A_1+2\sin\frac{2\pi x}{L})\epsilon 
\right.\nonumber \\
&&
\left.
+\left(\frac{-1}{8}(-A_1+2\sin\frac{2\pi x}{L})^2
-\frac{1}{2}A_2+
\frac{1}{2}\sin^2\frac{2\pi x}{L}+\frac{1}{8}B_1^2\right)\epsilon^2
+O(\epsilon^3)\right\}=L.\nonumber \\
&&
\end{eqnarray}
Noting that coefficients of each power of $\epsilon$ should vanish and 
using $\int_0^Ldx\sin\frac{2\pi x}{L}=0$, 
$\int_0^Ldx
\sin^2\frac{2\pi x}{L}=\frac{1}{2}$, we obtain
\begin{equation}
A_1=0, \;\;\;  A_2=\frac{1}{4}B_1^2.
\end{equation}
In the same way, the imaginary part of (\ref{perturbation}) is (provided $A_1=0$)
\begin{equation}
\int_0^L dx \left\{\frac{B_1}{2}\epsilon 
+\frac{1}{2}\left(B_2-B_1\sin \frac{2\pi x}{L}\right)\epsilon^2
+ O(\epsilon^3)\right\}
=\frac{2\theta L}{W}.
\end{equation}
Thus one has
\begin{equation}
A_2=\frac{4\theta^2}{U_0^2},\;\;\;  B_1=\pm \frac{4\theta}{U_0}.
\end{equation}
From these expressions the eigenvalue is given by Eq.(\ref{complexWKB}).
As shown in Fig.\ref{Figg3} this estimation of $E_1$ is very accurate for $W\gg U_0$.


\begin{thebibliography}{30}
\bibitem{Frenken}
J. W. M. Frenken and J. F. Van der Veen, Phys. Rev. Lett. {\bf54}, (1985) 134; B. Pluis et al., 
ibid. {\bf59} (1987) 2678 ; 
T. R. Linderoth, S.Horch, E. Laegsgaard, I. Stensgaard, and F. Besenbacher, ibid.{\bf78}
 (1997) 4978; P. Talkner, E. Hershkovitz, E. Pollak, and P. H\"anggi, Surf. Sci. {\bf437} 
 (1999) 198

\bibitem{Barone}
A. Barone and G. Paterno, {\em Physics and Applications of the Josephson Effect} (Wiley, New 
York, 1982)

\bibitem{Shapiro}
B. Shapiro,  M. Gitterman, I. Dayan, and G. H. Weiss, Phys. Rev B {\bf 46} (1992) 8416

\bibitem{Georgievskii}
A. Burshtein and Y. Georgievskii, J. Chem. Phys. {\bf 100}  (1994) 7319

\bibitem{Fulde}
P. Fulde, L. Pietronero, W. R. Schneider, and S. Str\"assler, 
Phys. Rev. Lett. {\bf35} (1975) 1776;
W. Dietrich, P. Fulde, and I. Peschel, Adv. Phys. {\bf29} (1980) 527

\bibitem{Gruner}
G. Gruner, A. Zawadowski, and P. M. Chaikin, Phys. Rev. Lett. {\bf46} (1981) 511

\bibitem{Ajdari}
A. Ajdari and J. Prost, Proc. Natl. Acad. Sci. U. S. A. {\bf88} (1991) 4468; G. I. Nixon 
and G. W. Slater, 
Phys. Rev. E {\bf53} (1996) 4969

\bibitem{Hoppenstead}
F. C. Hoppenstead and E. M. Izhikevich, {\em Weakly Connected Neural Networks} (Springer, 
New York, 1997);
K. Wiesenfeld, D. Pierson, E. Pantazelou, C. Dames, and F. Moss, 
Phys. Rev. Lett. {\bf72} (1994)  2125 ; 
C. Kurrer and K. Schulten, Phys. Rev. E {\bf51} (1995)  6213 

\bibitem{Reimann1}
P. Reimann, Phys. Rep. {\bf 361} (2002) 57 

\bibitem{Hangi}
P. H\"anggi, P. Talkner, and M. Borkovec, Rev. Mod. Phys. {\bf62} (1990) 251

\bibitem{Doering}
C. R. Doering, W. Horshtemke, and J. Riordan, Phys. Rev. Lett. {\bf 72} (1994) 2984 

\bibitem{Reimann2}
P. Reimann, C. V. Broeck, H. Linke, P. H\"anggi, J. M. Rubi, and 
A. P$\acute{e}$rez-Madrid, Phys. Rev. Lett. {\bf 87} (2001) 010602;
P.Reimann, C.V.Broeck, H.Linke, P. H\"angi, J.M.Rubi, and A.P$\acute{e}$rez-Madrid, 
Phys.Rev.E
{\bf65} (2002) 031104

\bibitem{Risken}
H. Risken, {\em The Fokker Planck Equation} ( Springer, Berlin, 1989)

\bibitem{Lebowitz}J. L. Lebowitz and P. G. Bergmann, Ann. Physik {\bf1} (1957) 1

\bibitem{vanKampen}
N. G. van Kampen, J. Stat. Phys. {\bf 17}  (1977) 71;
Supplement of Prog. Theor. Phys. {\bf 64} (1978) 389;
{\em Stochastic processes in physics and chemistry}, 2nd ed. 
(North-Holland, Amsterdam, 1981)

\bibitem{Caroli} B.Caroli,C.Caroli,and B.Roulet,
J. Stat. Phys. {\bf 21} (1979) 415;
B. Caroli, C. Caroli, and B. Roulet,
J. Stat. Phys. {\bf 26} (1981) 83;
H. Tomita, A. Ito,and H. Kidachi, Prog. Theor. Phys. {\bf 56} (1976) 786;
K. Nakamura and T. Sasada, Phys. Lett. A. {\bf 74} (1979) 379

\bibitem{Noziere} P. Nozi\'ere and G. Iche,
J. Physique {\bf 40} (1979) 225;
W. Dieterich {\it et al.}, Z. Physik B {\bf27} (1977) 177

\bibitem{threewell}
D ter Haar, {\em Selected Problems in Quantum Mechanics} (Academic, New York, 1964);
S. Ohta and K. Nakamura, J. Phys. C {\bf 14} (1981) L427

\end{thebibliography}
\end{document}